\documentclass[
    aps,
    prl,
    twocolumn,
    superscriptaddress
]{revtex4-2}

\usepackage[english]{babel}
\usepackage{amsmath,amssymb}
\usepackage{graphicx}
\usepackage{siunitx}
\usepackage{xspace}
\newcommand{\devC}{Device C\xspace}

\newcommand{\devE}{Device E\xspace}

\newcommand{\devJ}{Device J\xspace}
\newcommand{\pH}{p\mathrm{H}}
\DeclareSIUnit\Molar{\textsc{M}}
\newcommand{\hairsp}{\hspace{1pt}}
\newcommand{\ie}{\mbox{\textit{i.\hairsp{}e.}}\xspace}

\newcommand{\vs}{\mbox{\textit{vs.}}\xspace}
\newcommand{\tylabel}[1]{{\mbox{\tiny #1}}}
\begin{document}

\title{Electrochemical impedance spectroscopy of graphene nanogaps}

\author{Patrick A. McKee}
\author{Chris S. DeMellier}
\author{Robin N. Schipper}
\author{Henk W.Ch. Postma}
\email{postma@csun.edu}
\affiliation{Physics and Astronomy, California State University Northridge,
Northridge, California 91330, USA}

\begin{abstract}
Graphene nanogaps represent an emerging platform for nanoscale electrochemical and sensing devices, with potential applications in next-generation biomolecular sequencing. However, their interfacial behavior in aqueous environments remains poorly characterized, particularly with respect to frequency-dependent impedance and charge transport mechanisms at the graphene edge. We fabricate graphene nanogaps by controlled electrical breakdown in an inert atmosphere and study their electrochemical response. Upon exposure to ambient conditions, a surface contamination layer supports electrochemical activity within an adsorbed ultrathin conductive film between the graphene edges. Electrochemical impedance spectroscopy reveals distinct frequency-dependent responses consistent with a Warburg element associated with diffusion in this confined interfacial film. The impedance evolves systematically with liquid $\pH$, reflecting changes in electrochemical reaction-diffusion processes at the graphene edges. A quantitative equivalent-circuit model captures these effects and enables extraction of an effective nanogap length scale from impedance spectra, providing information complementary to that obtained from tunneling measurements.
\end{abstract}

\maketitle

\section*{Introduction}
Rapid sequencing of individual biomolecules promises to usher in the era of personalized medicine. Protein nanopores, such as those employed in Oxford Nanopore Technologies' platform, have gained popularity recently. However, the raw throughput speed is limited due to its necessary use of a ratcheting enzyme, and even with advanced neural networks, the MinION system struggles to tell apart similar sequences such as CCAGG and CCTGG, due to a confluence of poor base-calling accuracy and absence of true single-base resolution \cite{wick_performance_2019}. We have been developing a tunneling sequencing platform employing graphene nanogaps, which, due to its tunneling sensing mechanism, promises to resolve single bases at the full translocation speed of single molecules, many orders of magnitude faster than an ion-current sensing approach \cite{postma_rapid_2010}.

Graphene nanogaps can be formed by electromigration \cite{prins_room-temperature_2011,nef_high-yield_2014,puczkarski_graphene_2017}, following related work on carbon nanotube and metal junctions \cite{collins_current_2001,park_coulomb_2002,chiu_ballistic_2005}, and have also been realized in several alternative electrode geometries \cite{radha_molecular_2016,bellunato_dynamic_2018,arjmandi-tash_supramolecular_2020}. Planar graphene nanogaps are particularly attractive for biomolecular sensing because DNA can translocate through them \cite{patel_dna-graphene_2017} and trapped molecules can be electrically interrogated \cite{cao_building_2012,xin_stereoelectronic_2017,gu_building_2018,el_abbassi_robust_2019,yang_graphene-molecule-graphene_2023}. However, despite extensive work on nanogap fabrication \cite{prins_room-temperature_2011,standley_graphene-based_2008,lau_nanoscale_2014,nef_high-yield_2014,gu_building_2018,el_abbassi_robust_2019}, their electrochemical behavior in aqueous environments remains poorly understood.

Here, we make several-$\si{\micro\meter}$-long nanogaps in single-layer graphene at room temperature and study their electrochemical response in ambient conditions and liquid. We measure the nanogap formation time to be less than $\SI{1}{\micro\second}$; a detailed analysis of the formation process is beyond the scope of this work. We find that under ambient conditions and in liquid, the ever-present surface contamination layer leads to electrochemical processes that can be modeled by a thin liquid film on the surface between the graphene edges that constitutes the conductive medium, and a Warburg-like constant phase element that captures the reaction-diffusion dynamics. We find that the circuit elements can be changed by varying $\pH$. The effective nanogap length scale extracted from impedance differs from that obtained from tunneling measurements, allowing us to interrogate variations in nanogap width along the nanogap length.

\section*{Electrochemical impedance spectroscopy}
Graphene nanogaps are formed by patterning graphene into $0.5-\SI{4}{\micro\meter}$ wide constrictions using electron-beam lithography and reactive ion etching, followed by electrical breakdown at room temperature in an Ar atmosphere.

\begin{figure}[!h]
\includegraphics[width=\columnwidth]{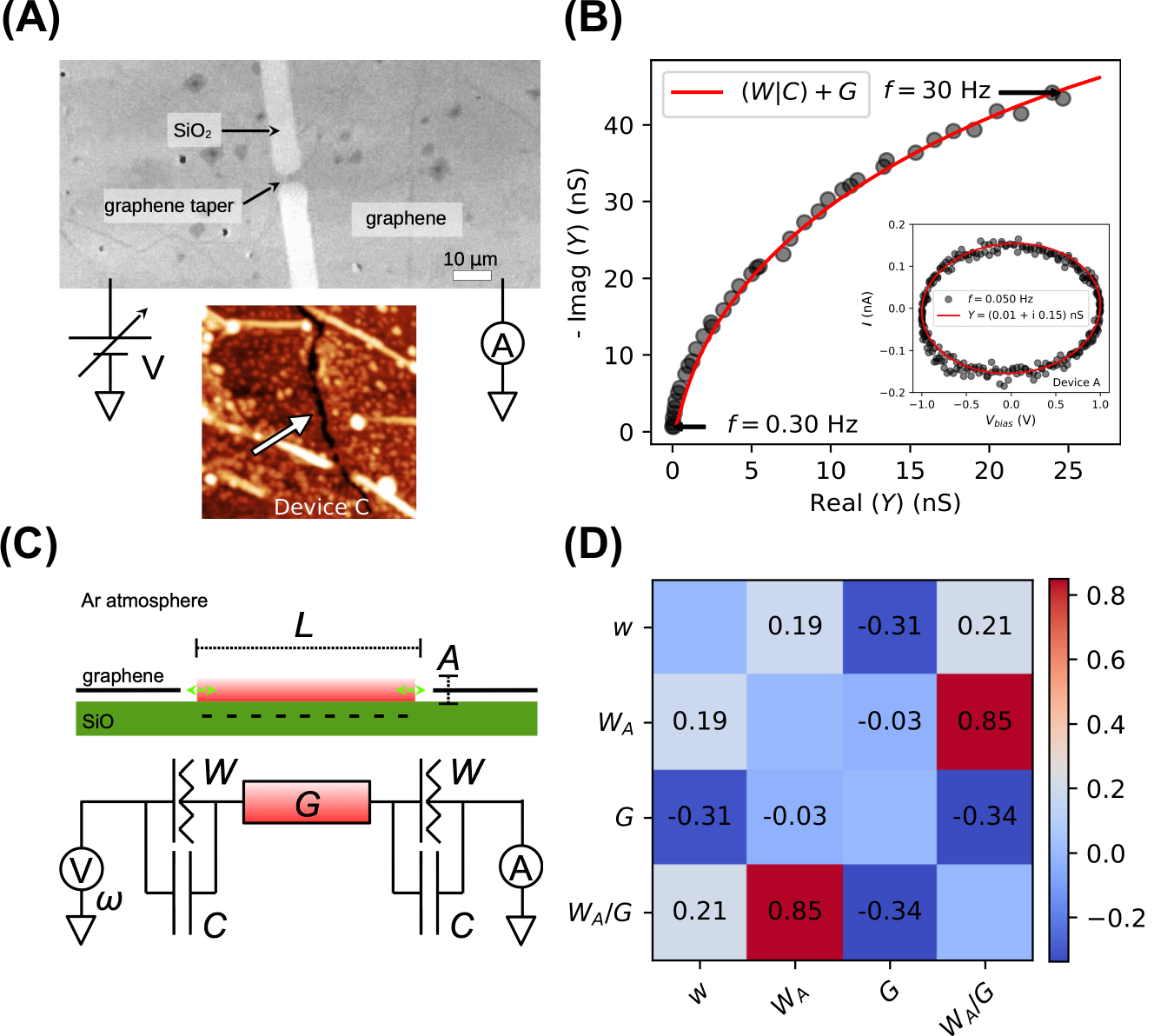}
\caption{\textbf{Graphene nanogaps exhibit a frequency-dependent electrochemical response that is captured by an equivalent-circuit model associated with a surface contamination layer.} (A, top) SEM image of a narrow patterned graphene constriction on a SiO$_2$ surface with taper width $w = \SI{3}{\micro\meter}$ driven by current $I$ sourced by a bias voltage $V$. (A, bottom) AFM height image of \devC after the break, with the nanogap indicated by the white arrow (image size $0.6\times\SI{0.6}{\micro\meter\squared}$). (B) Nyquist plot of $Y$ as a function of $f$ (black circles) and fit to the admittance model (red line). (Inset) $I$ \vs $V_{\tylabel{bias}}$ at $f=\SI{50}{\milli\hertz}$ (black points) and fit to $Y$ (red line). (C) Electrochemical conduction through the surface contamination layer (red) driven by the reaction with the graphene (green arrow) over a length $L$ and equivalent circuit with elements $W$, $C$, and $G$, driven by a source at $\omega$. (D) Pearson correlation coefficient $R$ between device parameters as indicated on horizontal and vertical axes of 16 devices without visible defects.}
\label{fign1}
\end{figure}

After the nanogap has formed, we sweep $V_\tylabel{bias}$ at angular frequency $\omega \equiv 2\pi f$ and record a device current $I(\omega)$. For the majority of the devices, we observe a slightly out-of-phase response (Fig~\ref{fign1}B). For the very smallest nanogaps we observe Simmons-like electron tunneling behavior \cite{simmons_generalized_1963,prins_room-temperature_2011,nef_high-yield_2014}, which we discuss later. To ensure we are in the linear response regime, we set the $V_\tylabel{bias}$ amplitude to $\SI{60}{\milli\volt}$ and vary $\omega$ to measure the admittance $Y(\omega) \equiv \frac{\partial I}{\partial V} (\omega) \equiv G(\omega) + \imath B(\omega)$, where $G$ and $B$ represent conductance and susceptance, resp. (Fig~\ref{fign1}B).

To identify an equivalent-circuit model for these responses, we considered a conductor $G$ with admittance $Y_G = G$, a capacitor $C$ with admittance $Y_C = \imath \omega C$, and a Warburg element with admittance $Y_W = \sqrt{\imath\omega}\, W_A$. Here $W_A \equiv 1/A_W$ is the inverse of the Warburg factor $A_W \equiv \frac{k_B T}{n^2 e^2 \sqrt{D} A C_0}$, with $k_B$ the Boltzmann constant, $T$ the absolute temperature, $n$ the valence, $e$ the elementary charge, $D$ the diffusion coefficient, $A$ the effective cross-sectional area of the conductive surface layer, and $C_0$ the equilibrium concentration. We fit the data to various topologies using these three components and find that the one with the lowest $\chi^2$ value is a capacitor in parallel with a Warburg element, in series with a conductor, \ie $(W | C) + G$ (Fig~\ref{fign1}C). $C$ is typically $\sim \SI{0.5}{\nano\farad}$ and is dominated by parasitic capacitive coupling between the leads, so we do not analyze it further. These measurements suggest that electrochemical processes are occurring within the graphene nanogap. As these measurements are performed in an inert Ar environment before explicitly adding liquid, we hypothesize that the electrochemical processes are due to the ubiquitous so-called `surface contamination layer'. This few-monolayer liquid layer is typically encountered on surfaces exposed to atmospheres with a partial pressure of vapors and can cause stiction in scanning probe microscopy or fabrication reproducibility issues \cite{macdonald_airborne_1993}. Although the nanogap region is locally heated during electrical breakdown \cite{nef_high-yield_2014}, the surrounding surface retains its contamination layer, which can quickly diffuse back into the nanogap \cite{pan_nanoconfined_2020}. A likely candidate is water, as that is commonly the solvent used for transferring graphene from its catalytic metal film to a target wafer. In addition, water has been shown to protect the graphene surface from further hydrocarbon contamination \cite{li_water_2016}. While water vapor can also sometimes originate in the Ar supply \cite{carroll_characterization_2019}, that is unlikely here because our Ar is fed through a desiccating filter.

The device characteristics are measured with relatively little electrical noise, and the fits closely follow the data. However, although we expect both $G$ and $W_A$ to be proportional to the effective cross-sectional area $A$, and therefore also to $w$, analysis of 16 devices shows very little correlation between the geometric properties and electrochemical circuit elements (Fig~\ref{fign1}D). We attribute these sample-to-sample variations to uncontrolled microscopic differences between devices. Therefore, a quantitative exploration of the circuit components requires holding these variations fixed while changing another parameter for the same device.

\begin{figure}[!h]
\includegraphics[width=\columnwidth]{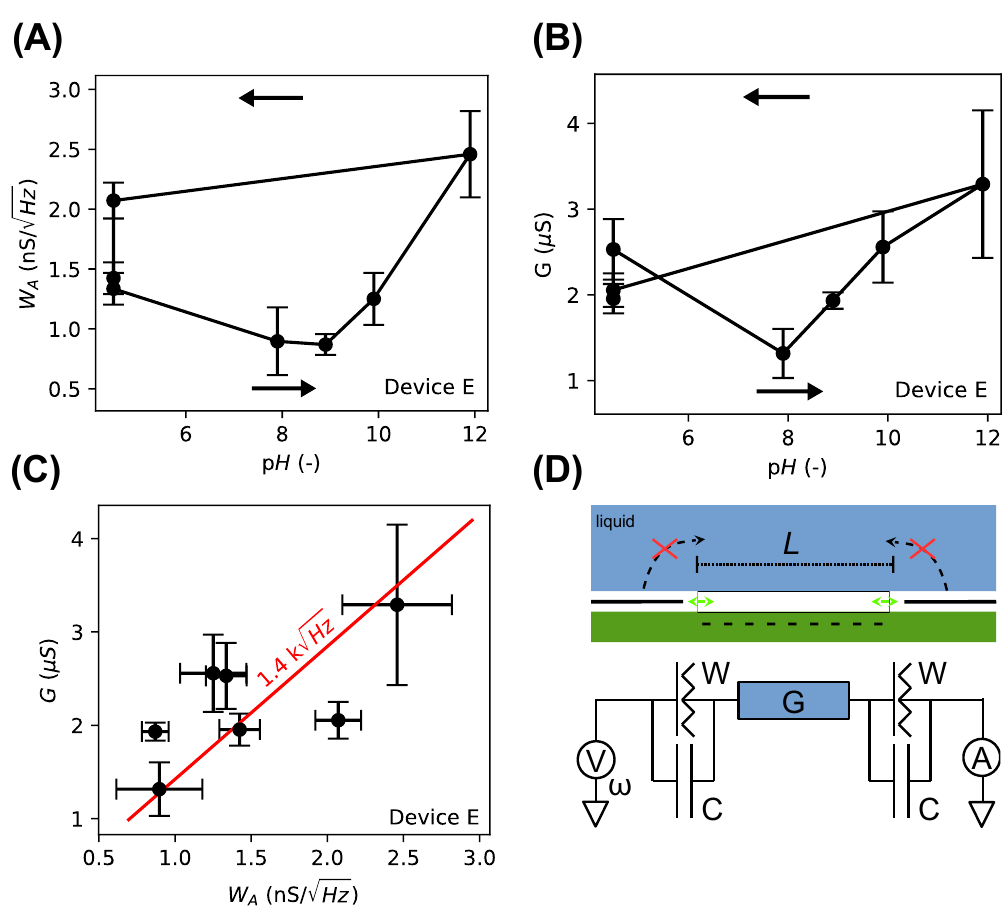}
\caption{\textbf{Dependence of $Y$ on $\pH$ for \devE.} (A) Dependence of $W_A$ on $\pH$. (B) Dependence of $G$ on $\pH$. (C) Linear correlation between $W_A$ and $G$ (black crosses) and linear fit (red line); error bars are parameter uncertainties reported by Lmfit. (D) Circuit diagram and microscopic diagram of the conduction process for a nanogap immersed in water solution (blue), with the conductance determined by the concentration close to the surface (red).}
\label{fign2}
\end{figure}

When we introduce water with varying $\pH$, we find relatively small changes in the circuit elements, despite the proton concentration changing by 7 orders of magnitude (Fig~\ref{fign2}). We find no systematic effect of introducing KCl with concentrations from $\SI{10}{\milli\Molar}$ to $\SI{2}{\Molar}$: $G$ varies by $< 2\%$ and $W_A$ by $<10\%$ (less than the error bars in Fig~\ref{fign2}C), and their variations are not correlated, unlike when changing $\pH$.

These observations are consistent with our hypothesis that electrochemical processes occur very close to the surface and between the graphene edges, due to the limited electrochemical activity of the graphene basal plane \cite{davies_nanotrench_2005} (red crosses, Fig~\ref{fign2}D). It is well known that the ionic concentration very close to a charged surface (in this case SiO$_2$) depends only weakly on $\pH$ \cite{smeets_salt_2006,schoch_transport_2008,bandara_conductance-based_2018}. We note that the experiments presented here are qualitatively different because the electrodes are much closer to the surface than the bulk of the liquid and therefore in a different transport regime.

There is a correlation between $G$ and $W_A$ for the same device (Fig~\ref{fign2}C), and we observe this on all 7 devices tested in this manner. For an Ohmic ionic thin conductor between the graphene edges over a nanogap distance $L$ (Fig~\ref{fign2}D), the conductance is $G = \sigma A/L$, with conductivity $\sigma = \mu_q C_0 e$, where $\mu_q$ is the charge mobility and $D = \mu_q k_B T/e$. Therefore,
\begin{equation}
\frac{G}{W_A} = \frac{\sqrt{D}}{L}.
\end{equation}
For the same device, $L$ and $D$ are fixed to first order, so $G/W_A$ is independent of the ionic concentration $C_0$. The model therefore predicts the observed correlation between $G$ and $W_A$ as $\pH$ is varied on the same device. In contrast, $L$ and other microscopic properties vary from device to device, consistent with the lack of correlation across devices noted above (Fig~\ref{fign1}D).

We assume the minimum in both $G$ and $W_A$ corresponds to the isoelectric point of the surface, whose precise measurement may yield detailed insight into the surface species \cite{bandara_conductance-based_2018}. From the ratio of the mean values, $\left\langle L \right\rangle /\left\langle W_A/G \right\rangle$, we find $D = \SI{1.1e-8}{\meter\squared\per\second}$. This value is rather low compared to typical values for ions in bulk water of $\sim \SI{8e-8}{\meter\squared\per\second}$ \cite{haynes_crc_2014}. It is not unexpected to find lower diffusion, as diffusion close to solid objects differs from its bulk value and is usually lower \cite{pan_nanoconfined_2020}.

We do not observe a strong linear correlation between $W_A/G$ and $L$ across devices, most likely due to microscopic sample-to-sample variations. For instance, isolated localized charges are often found on SiO$_2$ surfaces \cite{woodside_scanned_2002}, and they can cause variations in the thickness of the liquid surface layer. Further, competition between elastic bending and surface adhesion of the graphene changes the local height of the graphene electrode over the surface, and therefore the degree of contact with the surface contamination layer.

\section*{Simultaneous tunneling and impedance spectroscopy}
\begin{figure}[!h]
\includegraphics[width=\columnwidth]{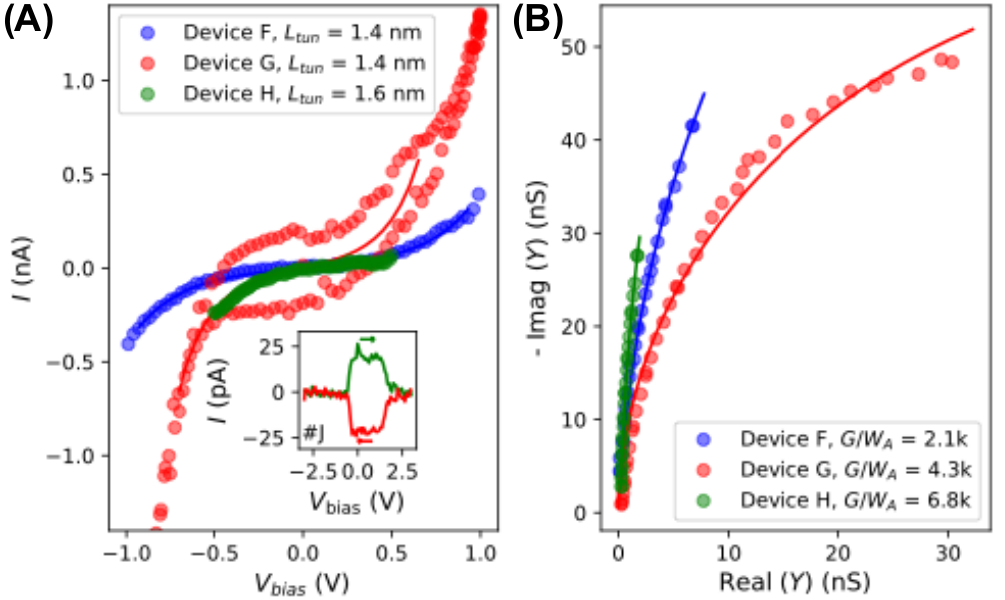}
\caption{\textbf{Tunneling curve and electrochemical admittance for the 3 devices with the smallest nanogap.} (A) High bias $I(V_\tylabel{bias})$ curves for the three indicated devices and fits to the Simmons model. (Inset to A) Up and down peaks in the up (green) and down (red) sweeps of \devJ, at a sweep rate of $\SI{0.5}{\volt\per\second}$. (B) Low bias $\omega$ dependence of $Y$ for the same three devices (circles) and fits to the impedance model $(W|C) + G$ and values of $G/W_A$ as indicated.}
\label{fign3}
\end{figure}

Most reported nanogap experiments use a tunneling analysis of the $I(V_\tylabel{bias})$ curve, using a variation of a Simmons model \cite{simmons_generalized_1963} to extract the nanogap geometry \cite{prins_room-temperature_2011,nef_high-yield_2014,lau_nanoscale_2014}. In our experiments we observe Simmons-like curves only for the smallest nanogaps fabricated, \ie smaller than $\sim \SI{1.6}{\nano\meter}$ (Fig~\ref{fign3}A), as for larger gaps the tunneling transport is electrically shorted by the electrochemical transport. We keep the voltage bias relatively low to prevent field-induced changes to the nanogap \cite{standley_graphene-based_2008}. For short nanogaps that show tunneling transport at high bias, we observe electrochemical transport at low bias (Fig~\ref{fign3}B). From the mean ratio $\left\langle L \right\rangle /\left\langle W_A/G \right\rangle$ we find $D = \SI{2e-11}{\meter\squared\per\second}$, much smaller than the value obtained for larger gaps above. We believe this apparent difference arises because, for irregular gaps, tunneling is dominated by the regions where the graphene edges are closest, at a distance $L_{\tylabel{tun}}$, whereas electrochemical transport occurs over a different effective distance $L \equiv L_{\tylabel{EIS}}$, where the gap is accessible to the liquid. A ratio of $L_{\tylabel{EIS}}/L_{\tylabel{tun}} \sim 22$ would be sufficient to recover the same $D$ as above. Simultaneous measurements of tunneling and electrochemical transport therefore yield insight into local variations in nanogap width. This information is difficult to obtain from tunneling alone and is also challenging to access by scanning probe microscopy at these length scales.

For a very small subset of devices, we observe peaks in the $I(V)$ curve that are reminiscent of cyclic-voltammetry oxidation-reduction traces (Fig~\ref{fign3}A inset). However, the individual electrode potentials and local ionic concentrations are not well controlled in the small space between the electrodes, and this is not readily mitigated with a conventional three-electrode setup. Further, a significant fraction of the applied potential may develop across the high-impedance surface contamination layer. These peaks are the subject of future investigations.

\section*{Materials and methods}
\subsection*{Device fabrication and nanogap formation}
Single-layer graphene grown by chemical vapor deposition was transferred to Si/SiO$_2$ substrates and patterned into $0.5-\SI{4}{\micro\meter}$ wide constrictions using electron-beam lithography and reactive-ion etching. A polymethyl methacrylate (PMMA) resist was patterned by electron-beam lithography, followed by development in methyl isobutyl ketone and isopropyl alcohol (1:3) and oxygen-plasma etching. The PMMA was subsequently removed in acetone, except for devices used in the $\pH$ measurements described below.

Devices were mounted in a custom flow cell providing electrical access and gas or liquid access to the graphene surface. Nanogaps were formed at room temperature in an Ar atmosphere by slowly increasing the applied voltage until an abrupt decrease in current indicated electrical breakdown of the graphene constriction. For constriction widths of $0.5-\SI{4}{\micro\meter}$, the corresponding breaking currents ranged from $0.5 - \SI{5.0}{\milli\ampere}$. The breaking event was monitored using a high-bandwidth electrical readout and was measured to occur in less than $\SI{1}{\micro\second}$. A detailed analysis of the dependence of the breaking current and dynamics on device geometry is beyond the scope of this work.

\subsection*{Electrochemical impedance measurements}
Following nanogap formation, the high-bias drive circuitry used for electrical breakdown was disconnected and the devices were measured at low bias using a current amplifier. For impedance measurements, the applied voltage peak amplitude was reduced to $\SI{60}{\milli\volt}$ to remain in the linear-response regime. The device current was recorded while varying the angular frequency $\omega=2\pi f$, and the complex admittance was calculated as
\begin{equation}
Y(\omega)=\frac{\partial I}{\partial V}=G(\omega)+\imath B(\omega),
\end{equation}
where $G$ and $B$ are the conductance and susceptance, respectively.

An initial coarse frequency sweep was used to determine the frequency range containing the dominant response, after which additional frequencies were selected to sample the Nyquist response more uniformly in the $(\mathrm{Real} (Y), \mathrm{Imag} (Y))$ plane. For every frequency point, more than one cycle was recorded and the first cycle was discarded to eliminate transients at the start. 
Initial measurements were extended to frequencies as high as $\SI{50}{\kilo\hertz}$ to test whether additional circuit elements contributed to the response. Once no additional high-frequency features were observed, subsequent measurements were restricted to frequencies below $\SI{100}{\hertz}$.

The admittance spectra were fit to equivalent-circuit models constructed from a conductor, capacitor, and Warburg element, with
\begin{equation}
Y_G=G,\qquad
Y_C=\imath\omega C,\qquad
Y_W=W_A\sqrt{\imath\omega}.
\end{equation}
All distinct circuit topologies containing these elements were compared, and the topology $(W|C)+G$ produced the lowest $\chi^2$ values for the data presented here. Fits were performed in Python using the Lmfit package, and parameter uncertainties were obtained from the fitting procedure.

\subsection*{pH and electrolyte measurements}
For $\pH$ measurements, devices were covered with PMMA before nanogap formation. Electrical breakdown locally melted or ablated the polymer near the constriction, providing liquid access primarily to the nanogap and its immediate surroundings. The $\pH$ was adjusted by titrating KOH into filtered deionized water. Following each solution exchange, the electrical response was monitored until a stable value was reached before the impedance spectrum was recorded.

The dependence of the fitted circuit parameters on $\pH$ was measured on seven devices. As a control for the effect of bulk ionic concentration, KCl solutions with concentrations ranging from $\SI{10}{\milli\Molar}$ to $\SI{2}{\Molar}$ were also introduced.

Control measurements were additionally performed on open-face devices without a PMMA covering or ablated opening, for which a drop of liquid could be applied and completely removed. These devices were optically inspected following liquid exchange to verify that air was not trapped at the nanogap.

\subsection*{Imaging and device selection}
Nanogaps were characterized by atomic force microscopy (AFM) or scanning electron microscopy (SEM). Because SEM imaging can modify the electrical conductance of graphene, most devices were imaged after electrical measurements rather than before them.

Electrochemical admittance measurements were performed on 94 devices. The cross-device correlation analysis in Fig~\ref{fign1}D was restricted to 16 devices without visible structural defects. The $\pH$-dependent analysis was performed on seven devices for which repeated liquid exchange and stable impedance measurements could be obtained. Microscopic differences in the graphene--substrate interface and surface contamination layer were not otherwise used as selection criteria.

\section*{Conclusions}
We show that under ambient conditions, graphene nanogaps exhibit electrochemical behavior consistent with a conductive thin liquid film between the graphene edges and a Warburg-like element. This model allows an effective nanogap length scale to be estimated from the circuit element values and yields information complementary to that obtained from tunneling measurements. We further find that varying $\pH$ produces correlated changes in $G$ and $W_A$ on the same device, consistent with the relation $G/W_A=\sqrt{D}/L$. Going forward, cyclic-voltammetry studies of the surface contamination layer may help identify the species and functional groups at the graphene edge that dominate the electrochemistry in the nanogaps. These results provide a basis for understanding and ultimately operating graphene nanogaps in aqueous sensing and sequencing environments.

\section*{Acknowledgments}
We thank Konstantin Daskalov, Aracely Gutierrez, Kaden Kalstrom, Ravipa Losakul, Julia Postma, and Alex Stafford for experimental assistance and Jason Dwyer, David Hoogerheide, Jay Kim, Jinyang Li, Miroslav Peric, and Derek Stein for discussions.

\bibliography{refs}
\end{document}